# On the probability of habitable planets.

**François Forget**

LMD, Institut Pierre Simon Laplace, CNRS, UPMC, Paris, France

E-mail : forget@lmd.jussieu.fr

## Abstract

In the past 15 years, astronomers have revealed that a significant fraction of the stars should harbor planets and that it is likely that terrestrial planets are abundant in our galaxy. Among these planets, how many are habitable, i.e. suitable for life and its evolution? These questions have been discussed for years and we are slowly making progress. Liquid water remains the key criterion for habitability. It can exist in the interior of a variety of planetary bodies, but it is usually assumed that liquid water at the surface interacting with rocks and light is necessary for the emergence of a life able to modify its environment and evolve. A first key issue is thus to understand the climatic conditions allowing surface liquid water assuming a suitable atmosphere. This have been studied with global mean 1D models which has defined the "classical habitable zone", the range of orbital distances within which worlds can maintain liquid water on their surfaces (Kasting et al. 1993). A new generation of 3D climate models based on universal equations and tested on bodies in the solar system is now available to explore with accuracy climate regimes that could locally allow liquid water.

A second key issue is now to better understand the processes which control the composition and the evolution of the atmospheres of exoplanets, and in particular the geophysical feedbacks that seems to be necessary to maintain a continuously habitable climate. From that point of view, it is not impossible that the Earth's case may be special and uncommon.

# 1 Introduction

The recent detections of many extrasolar planets have allowed us to make a major step forward in our investigation of the ultimate question in astrobiology: "are we alone ?". In particular, we will soon be able to estimate one of the first terms of the Drake equation, the fraction of stars with planets, and even evaluate the abundance of planets of a given size (including rocky or ocean planets) in a given range of orbital distance from the different type of stars. Already, meaningful statistics are available, notably from the hundreds of planets detected by the radial velocity method using Earth-based telescopes (e.g. Mayor and Queloz, 2012, Howard et al. 2010), and from the thousands (>2600 in 2013) of planetary candidates observed to transit in front of their star by the Kepler space telescope (see e.g. Borucki et al. 2011). These are called "candidates" because a fraction may be artifacts, for instance due to eclipsing binary stars in the background of or within the observed system. This can mimic the photometric signal of a transiting planet.. Theoretical studies suggest a "false positive rate" of about 10% on average, up to 18% for giant planets (Fressin et al. 2013), although recent radial velocity observations of Kepler candidates suggest a false positive rate higher than 30% for hot jupiters (Santerne et al. 2012).

Nevertheless, to first order, the evaluation of the distribution of planets detected by radial velocity or surveys of transits should be meaningful. Of course, both methods are biased because the detection of small planets (i.e. Earth-size) remains very difficult, especially if they have long orbital periods and if they orbit stars more massive than red M dwarf stars. Nevertheless, we have already learned a lot from the detection of "super-Earth", "small Neptune" and gas giants. Another method, based on gravitational lensing (Beaulieu et al. 2006, Cassan et al. 2012) provides unbiased statistics on a much smaller population, and confirm the general conclusions : planets, and in particular planets small enough to have a solid or liquid surface should be abundant in our galaxy. For instance, on the basis of radial velocity data and a simple extrapolation of the planetary distribution, Howard et al. (2010) predicted that "23% of stars harbor a close-in Earth-mass planet (ranging from 0.5 to 2.0 Earth masses)" and Bonfils et al. (2013) calculated that about 50% or more M dwarf stars should have terrestrial planets (>1-10 Earth masses) with orbit period between 10 and 100 days. Similarly, using only Kepler's first 4 months of data, Borucki et al. (2011) estimated that the frequency of stars observed by Kepler harboring planets with a diameter lower than twice the Earth and a short period lower than 50 days is 13%. It is likely that the actual percentage of stars with terrestrial planets should be much higher. A few interesting systems with potentially rocky planet candidates (mass < 10 Earth masses) close to the habitable zone have already been discovered: Gliese 581 (Udry et al. 2007; Mayor et al. 2009), HD85512b (Pepe et al. 2011), Gliese 667Cc (Bonfils et al. 2013), HD 40307g (Tuomi et al. 2013), Kepler-22b (Borucki et al. 2012).

Among these numerous planets, which fraction may be suitable for life to start and evolve? In this short review, written for non-specialists, I discuss the debates and scientific investigations relevant to this question. A particular emphasis is put on the long-term habitability which is probably necessary for life to 1) modify its environment so that we can detect it remotely (e.g. Kaltenegger et al. 2007) and 2) evolve to reach a technological level allowing the use of radio-signal which could be detected from the Earth, as estimated in the Drake equation.

In section 2, I introduce the classical notion of habitability, as defined by the occurrence of surface liquid water. In section 3, I discuss how climatological studies can constrain the range of planets which may be habitable, assuming that they enjoy a favorable atmosphere. Finally, in section 4, I show that the processes which have allowed the Earth to keep such a favorable atmosphere for billions of years are still poorly known, and that from that point of view, the Earth's case may be special and uncommon.

## 2 Habitability and surface liquid water

### 2.1 What makes a planet suitable for life?

With only our planet as a viable example, and only one kind of life to define the necessary ingredients, addressing this question requires enormous scientific extrapolations and some trust in purely theoretical studies.

Obviously, the answer depends on the kind of life that we want to consider. Life as we know it always uses carbon-based molecules with liquid water as a solvent, with no exceptions. In fact, our experience on Earth has told us that the requirement for life is liquid water, regardless of mean temperature and pressure (Brack, 1993). Living organisms can exist and thrive in almost any conditions on Earth if liquid water is available (Rothschild and Mancinelli 2001). Conversely, no creatures can "live" (i.e., have metabolic activity) without liquid water. One can speculate on forms of life based, say, on liquid ammonia, condensed methane or even plasma ions interactions. However, exploring the wide field of modern chemistry and challenging the most open-minded chemists reveals that with our present knowledge it is difficult to imagine any alternative chemistry approaching the combination of diversity, versatility and rapidity afforded by liquid water-based biochemistry. This results from the unique ability of carbon to form complex species, and the unique characteristics of water as a liquid solvent (a large dipole moment, the capability to form hydrogen bonds, to stabilize macromolecules, to orient hydrophobic-hydrophilic molecules, etc.).

Carbon is common in our galaxy. A surprisingly high number of molecules that are used in contemporary biochemistry on the Earth (including amino acids) are found in the interstellar medium, planetary atmospheres and surfaces, comets, asteroids and meteorites and interplanetary dust particles. (e.g. Henning and Salama, 1998, Ehrenfreund et al. 2011)Within this context, the primary definition of habitability is the presence of liquid water. This may be narrow-minded, but if optimistic conclusions can be reached with such a focus, then whatever we have ignored will only serve to broaden the biological arena (Sagan 1996). On the other hand, it can be argued that liquid water by itself may not be sufficient and that a few other elements and a source of energy (chemical gradient or light) are necessary to support life forms. However, the discovery of a very large variety of extremophiles on Earth in recent years suggests that just about every chemical gradient imaginable can support some sort of life (Lammer et al. 2009). The apparent rapid emergence of life (which seems to have existed on the Earth as far back as we can look) also tend to suggest that life may be common once a suitable liquid water habitat is available.

Water is abundant in our galaxy (e.g., Cernicharo and Crovisier, 2005) and is expected to be part of the initial inventory of terrestrial planets.

In practice, the primary difficulty to get liquid water is thus to be in the right range of temperature and pressure. Pressures must be significantly higher than the triple point (near 6.1 mb). Temperatures should range between the freezing point (0°C, or lower with dissolved salts) and the boiling point, which depends on the pressure.

For life to evolve from simple bacterial life into complex forms, about three billions of years have been necessary on our planet. I will not discuss here the inherent difficulty of biological evolution and the fact that "a lot of luck" may be required to make animals (Ward and Brownlee, 2000, Carter, 2008). Nevertheless, it is striking that when assessing the odds of having planets harboring complex or even intelligent life like in the Drake equation, one must estimate the frequency of planets which can remain continuously habitable for billions of years.

## 2.2 Four classes of habitable planets

In their review of the factors which are important for the evolution of habitable Earth-like planets, Lammer et al. (2009) proposed a classification of four liquid-water habitat types which I find very useful to structure the scientific debate on habitability. I propose here a slightly simplified version of these classes:

**Class I** habitats represent Earth-like analog planets where stellar and geophysical conditions allow water to be available at the surface, along with sunlight. Light is important because the most productive natural way of powering an organism is by either using sunlight via photosynthesis, or by digesting something that does (at least for life as we know it). On Earth, even most subsurface ecosystems derive their energy from photosynthesis. The deep-sea vent communities derive energy from the reaction of $H_2S$ from the vent with $O_2$ from the ambient seawater. However the $O_2$ comes from surface photosynthesis, so these ecosystems are ultimately also dependent on it. Only three ecosystems completely independent of photosynthesis have been reported, all of which have limited metabolisms (McKay et al. 2008).

**Class II** habitats include bodies which initially enjoy Earth-like conditions, but do not keep their ability to sustain liquid water on their surface due to stellar or geophysical conditions. Mars, and possibly Venus are example of this class. On such planets, it is reasonable to assume that life could start, and that this life could potentially migrate to the limited habitats left once the planet is no longer able to sustain liquid water on its surface. On Mars, for instance, deep subsurface aquifers are considered to be potential sites for remnant life, while on Venus it is speculated that some exotic lifeforms could be present in the liquid cloud droplets of the upper atmosphere.

**Class III** habitats are planetary bodies where water oceans exist below the surface, and where the oceans can interact directly with a silicate-rich core. Such a situation can be expected on water-rich planets located too far from their star to allow surface liquid water, but on which subsurface water is in liquid form because of the geothermal heat. An example of such an environment is given by Europa, one of Jupiter's satellites, which has only about a hundredth of Earth's mass and almost no atmosphere, but which is strongly heated by internal deformation resulting from tidal forces. In such worlds, not only is light not available as an energy source, but the organic material brought by meteorites (thought to have been necessary to start life in some scenarios) may not easily reach the liquid water. Nevertheless, interaction with silicates and hydrothermal activity, also thought to be important for the origin of life, are possible.

**Class IV** habitats are very water-rich world which have liquid water oceans or reservoirs lying above a solid ice layer. Indeed, even if most planets are expected to possess a silicate core covered by a water layer, if this layer is thick enough, water at its base will be in solid phase (ice polymorphs) because of the high pressure. Ganymede, Callisto are likely examples of this class. Their oceans are thought to be enclosed between thick ice layers. In such conditions, the emergence of life may be very difficult because the necessary ingredient for life will be likely completely diluted. Lammer et al. (2009) considered the lack of rocky substrate such a severe constraint that they also put the "Ocean planets" with the ocean lying over a thick ice layer in Class IV, even if the water was liquid at the surface and therefore exposed to light and meteoritic inputs.

Considering these four classifications, it seems hard to imagine higher life forms as we know them populating anything but a Class I habitable planet (Lammer et al., 2009). Furthermore, if a planet can only harbor life below its surface, the biosphere would not likely modify the whole planetary environment in an observable way (Rosing 2005), or even build Radiotelescopes. Detecting its presence on an exoplanet would thus be extremely difficult.

Within that context, the concept of *habitable* exoplanets is usually limited to surface habitability, and the term "habitable zone" is usually defined as the range of orbital distances within which worlds can maintain liquid water on their surfaces.

# 3    Being at the right distance from a star.

## 3.1    The classical habitable zone

The key reference on the estimation of the habitable zone remains the masterly work of Kasting, Whitmire and Reynolds (1993) (see references therein for previous studies). An up-to-date description of the habitable zone is also available from Selsis et al. (2007) in the framework of their assessment of the habitability of the planets around the star Gliese 581.

Most studies which still define the "classical habitable zone" are based on 1D climate modeling, which assesses the habitability of an entire planet by calculating the global average conditions using a single atmospheric column illuminated by the global averaged flux.

## 3.2    Inner edge of the habitable zone.

The classical inner edge of the habitable zone is the distance where surface water is completely vaporized or where the warm atmospheric conditions allows water to reach the upper atmosphere. There, it can be rapidly dissociated by ultraviolet radiation, with the hydrogen lost to space (the Earth currently keeps its water thanks to the cold-trapping of water at the tropopause). As thus defined, the inner edge may not be very far inside Earth's current orbit because of a destabilizing mechanism called the "runaway" greenhouse effect: if a planet with liquid water on its surface is "moved" toward the sun, its surface warms, increasing the amount of water vapor in the atmosphere. This water vapor strongly enhances the greenhouse effect, which tends to further warm the surface. On the basis of simple 1D model calculations, Kasting (1988) found that on an Earth-like planet around the Sun, oceans would completely vaporize at 0.84 Astronomical Units (AU). However, he also showed that the stratosphere would become completely saturated by water vapor at only 0.95 AU, quickly leading to the loss of all water. Clearly, this "water-loss" limit is the one of primary physical concern on the inner edge of the habitable zone. A lot of uncertainties exist, and the 0.95 AU limit can be considered to be conservative, mostly because clouds feedbacks are ignored (Kasting et al. 1993). Assuming that clouds may protect a planet by raising its albedo up to 80% (this approximately corresponds to continuous and thick water cloud cover), a habitable planet at about 0.5 AU from the sun is conceivable. This is an extreme value: physical processes able to maintain liquid water at say, 0.4 AU, are hard to imagine.

## 3.3    Outer edge of the habitable zone.

The classical outer edge of the habitable zone is the limit outside which water is completely frozen on the planet surface. Estimating this limit with a classical model of Earth climate with a present-day atmosphere would suggest that this limit is very close to the current Earth orbit because of strong positive feedbacks on the temperature related to the process of "runaway glaciation": a lower solar flux decreases the surface temperatures, and thus increases the snow and ice cover, leading to higher surface albedos which tend to further decrease the surface temperature (Sellers, 1969; Gérard et al. 1992; Longdoz and François 1997).

In reality, on Earth there is a long-term stabilization of the surface temperature and $CO_2$ level due to the carbonate-silicate cycle (Walker et al. 1981). This may be the case on other planets, assuming that they are geologically active and continuously outgas or recycle $CO_2$, and that carbonates form in the presence of surface liquid water. Consequently, $CO_2$ accumulates until the geological source is balanced by the liquid water sink, which ensures the presence of liquid water (this key assumption is discussed in Section 4.). Within this context, one can define the outer edge of the habitable zone as the limit where a realistic atmosphere - in terms of composition and thermal structure - can keep its surface warm enough for liquid water. The most likely greenhouse gases on an habitable planet are $CO_2$ and of course $H_2O$. Other gases like $NH_3$ or $CH_4$ are possible in a reducing atmosphere, but they are rapidly photodissociated so that they must be shielded from solar UV (Sagan and Chyba 1997) or produced by a continuous source or a recycling process (Kasting 1997). It turns out that a thick $CO_2$ atmosphere may be one of the most efficient solutions for keeping a planet warm. This is not only due to the properties of the $CO_2$ gas itself. In fact, the greenhouse effect of a purely gaseous atmosphere is limited, and in particular adding more and more greenhouse gas to keep a planet warm does not work indefinitely. The infrared opacity tends to saturate, while the absorbed solar energy decrease because of the increase of the albedo by Rayleigh scattering. If we consider a cloud-free $CO_2$ atmosphere (with a water pressure fixed by temperature), the classical habitable zone outer edge is at 1.67 AU from the present Sun (with a $CO_2$ pressure of about 8 bar; Kasting et al., 1993). Recent work suggests that this value is too large, because the $CO_2$ gas opacity was probably overestimated in Kasting et al. (1993)'s model (see Wordsworth et al., 2010a). However, taking into account the radiative effects of the $CO_2$ ice clouds, which tend to form in such thick $CO_2$ atmospheres allows further increases in the warming of the surface thanks to a cloud "scattering greenhouse effect" (Forget and Pierrehumbert, 1997). Taking into account this process, the outer edge of the habitable zone has been extended as far as 2.5 AU. However, more realistic simulations that spatially resolve the formation and effects of the clouds are required to confirm this value. For now, the value of 2.5 AU may be considered an optimistic upper limit for planets ressembling the Earth; i.e. with an atmosphere mainly composed of $CO_2$, $N_2$, $H_2O$ atmosphere.

Recently, Pierrehumbert and Gaidos (2011) explored the possible conditions on a very different kind of planets: super Earths that would have been able to retain an appropriate part of the primordial $H_2$-He mixtures which had accumulated during their formation. They showed that the spectroscopic process of "collision-induced absorption" allows molecular hydrogen to act as an incondensible greenhouse gas and that bars or tens of bars of pure $H_2$ could maintain surface temperatures above the freezing point of water well beyond the "classical" habitable zone defined for $CO_2$ greenhouse atmospheres, out to 10 AU from a solar-type star. A problem with this scenario is that $H_2$-rich envelope tends to either quickly escape to space after the planet formation (see section 4.1)., or stay very thick, preventing water from being liquid by keeping the surface pressure too high. Therefore, the percentage of planets left with exactly the right atmospheric pressures to allow habitable temperatures after the early stages of atmospheric erosion is likely to be very small (Wordsworth, 2012). Nevertheless, Wordsworth (2012) noted that the numerous Exoplanets well outside the "classical" habitable zone and indeed losing their primordial atmosphere would pass through transient periods where oceans could form on their surfaces. However, the duration of these habitable conditions would range from thousands to a few millions of years

### 3.4 Around other stars

To first order, the limits given above for the solar system can be extrapolated to planets orbiting other stars by scaling the orbital distance to the same stellar luminosity, which strongly depends on the stellar mass (Fig. 1). However, **stars smaller that the Sun** with low effective temperature emit their peak radiation at longer wavelengths (red and near-infrared), where the radiation is less reflected by the atmospheric Rayleigh scattering and a water rich atmosphere is more absorbant. In such conditions the planet is more efficiently heated. The edges of the habitable zone are shifted accordingly (Kasting et al. 1993 ; Fig 1). In fact, small M stars with masses 0.1-0.5 times the mass of the sun are particularly interesting, because they constitute approximately 75 % of the stellar population in our Galaxy, and have a negligible evolution in 10 Gyr. Thus, their "continuously habitable zone" is identical to their initial habitable zone. Terrestrial planets around M stars are also easier to detect! However, estimating their habitability requires us to address several exotic problems such as tidal resonance / locking (in the extreme 1:1 case, this means that one side of the planet will always face the star), active stellar flarings, and the related atmospheric escape, as discussed below (See Tarter et al. 2007, Buccino et al. 2007, Joshi 2003, Selsis et al. 2007).

**Stars larger than the Sun** are much less numerous in the Galaxy, and have a shorter lifetime. If one assumes that, say, 2 Gyr are necessary for complex organisms to form (and build radiotelescopes), only stars with masses less than 1.5 solar mass can be considered. A large stellar mass also affects the radiation output from the star, which is emitted at shorter wavelengths (blue and ultraviolet). The stellar light is more readily reflected by an atmosphere, and the habitable zone is shifted accordingly.

### 3.5  Re-exploring the habitability using 3D climate models

Before 2011, nearly all studies of habitability have been performed with simple 1D steady-state radiative convective models that simulate the global mean conditions. Exceptions to this rule have either been parameterised energy-balance models (EBMs) that study the change in surface temperature with latitude only (Williams and Kasting 1997, Spiegel et al. 2008), or three-dimensional simulations with Earth climate models (Joshi 2003).

In many cases, 1D models may not be sufficient to estimate the habitability of a planet. The next step in this investigation is thus to use 3D time-marching global climate models (GCMs) which are necessary to really understand the habitability of a planet: First, they allow simulation of *local* habitability conditions due to e.g., the diurnal and seasonal cycles, which lets us investigate the meaning of the habitable zone more precisely than is possible in a globally averaged simulation. They also help us to better understand the distribution and impact of clouds, which are of central importance to both the inner and outer edges of the habitable zone, as discussed earlier. Finally, 3D models allow predictions of the poleward and/or nightside transport of energy by the atmosphere and, in principle, the oceans. This is necessary to assess if the planetary water or a $CO_2$ atmosphere will collapse on the night side of a tidally locked planet, or at the poles of a planet with low obliquity.

In our Solar system, the climate of most planets can now be predicted using such global climate models (GCMs). In fact, a full GCM can be considered as a "planet simulator" that aims to simulate the complete environment on the basis of universal equations only. These models have initially been developed for Earth, as atmosphere numerical weather prediction models (designed to predict the weather a few days in advance) and global climate models (designed to fully simulate the climate system and its long term evolution). Such models are now used for countless applications, including tracer transport, coupling with the oceans or the geological $CO_2$ cycles, photochemistry, data assimilation to build data-derived climate database, etc... Because they are almost entirely built on physical equations (rather than empirical parameters), several teams around the world have been able to succesfully adapt them to other terrestrial planets or satellites. For instance, our team at Laboratoire de Meteorologie Dynamique has adapted the "LMDZ" Earth model to Mars (Hourdin et al. 1993, Forget et al. 1999), Titan (Hourdin et al. 1995, Lebonnois et al. 2012), Venus (Lebonnois et al. 2009) and soon Triton and Pluto. These models are used to predict and simulate the volatile cycles, atmospheric photochemistry, clouds and aerosols, past climates, and more.

We have recently developed a new type of climate model flexible enough to simulate the wide range of conditions that may exist on terrestrial exoplanets, including any atmospheric cocktail of gases, clouds and aerosols for any planetary size, and around any star. In practice, GCMs simulate (a) the motion of the atmosphere, including heat and tracer transport on the basis of the equations of hydrodynamics (b) the heating and cooling of the atmosphere and surface by solar and thermal radiation (i.e., the radiative transfer) (c) The storage and diffusion of heat in the subsurface (d) The mixing of subgrid-scale turbulence and convection and (e) the formation, transport and radiative effects of any clouds and aerosols that may be present. Additional levels of complexity may include ice formation / sublimation, interaction with oceans, and even the effects of vegetation and the biosphere. a) c) and d) are almost universal processes. We have learned from studying the solar system that the corresponding parameterizations can be applied without changes to most terrestrial planets. The radiative transfer equations are also universals, but to simulate the 3D climates on a new planet, one challenge has been to develop a radiative transfer code fast enough for 3D simulations and versatile enough to model any atmospheric cocktail or thick atmosphere accurately

This model is now applied to better understand the limit of the habitable zone. For instance, Wordsworth et al. (2011) applied it to explore the habitability of planet Gliese 581d, discovered in 2007 (Udry et al. 2007). Gliese 581d receives 35% less stellar energy than Mars and is probably locked in tidal resonance, with extremely low insolation at the poles and possibly a permanent night side. Under such conditions, it was unknown whether any habitable climate on the planet would be able to withstand global glaciation and / or atmospheric collapse. 1D models could not be conclusive (Wordsworth et al.2010b; von Paris et al. 2010, 2011, Kaltenegger et al. 2011). Wordsworth et al. (2011) performed three-dimensional climate simulations that demonstrated that Gliese 581d would have a stable atmosphere and surface liquid water for a wide range of plausible cases, making it the first confirmed super-Earth (exoplanet of 2-10 Earth masses) in the habitable zone. Taking into account the formation of $CO_2$ and water ice clouds, they found that atmospheres with over 10 bar $CO_2$ and varying amounts of background gas (e.g., $N_2$) yield global mean temperatures above 0°C for both land and ocean-covered surfaces (Figure 2).

Similarly, Leconte et al. (2013) have applied the 3D LMD Global Climate model to explore the possible climate on warm tidally locked planets such as Gliese 581c and HD85512b. With the same stellar flux, a planet like the Earth would not be habitable because of the runaway greenhouse instability (see section3.2). On a tidally locked planet, they found that two stable climate regimes could exist. One is the classical runaway state, and the other is a collapsed state where water is captured in permanent cold traps (i.e. on the night side). If a thick ice cap can accumulate theret, gravity driven ice flows and geothermal flux should come into play to produce long-lived liquid water at the edge and/or bottom of the ice cap, well inside the inner edge of the classical habitable zone.

## 4   Having the right atmosphere.

Staying in the habitable zone is obviously not sufficient for a planet to continuously maintain liquid water on its surface: it must have an atmosphere which keeps the surface pressure and the surface temperature (through its greenhouse effect) in the right range, for billions of years. However, the processes which controls the atmospheric evolution on a planet are still poorly known. This is the major source of uncertainty regarding the probability of habitable planets. Below I briefly discuss two examples of processes (among many others) for which the Earth's case may be special rather than universal.

## 4.1 Habitability and atmospheric escape.

The first process which governs the long term evolution of an atmosphere is the atmospheric escape to space. To first order, it depends on the gravity, and on the temperature of the upper layer of the atmosphere (the exobase) where the atmospheric molecules can escape the planet's gravity field if they are fast enough, i.e. if the exobase temperature is high enough. (note that, in some cases, escape can also result from chemical reactions or interactions with the stellar wind). The temperature of the exobase is not controlled by the total insolation which heats the surface and the lower atmosphere. Instead, it depends on the flux of energetic radiations and the plasma flow from the star (especially the extreme ultraviolet which is absorbed by the upper atmosphere). It is also controlled by the ability of the atmospheric molecules to radiatively cool to space by emitting infrared radiation. To simplify, greenhouse gases like $CO_2$ can efficiently cool, whereas other gases like $N_2$ cannot.

To keep its atmosphere and remain habitable, a planet must be large enough and exert a high enough gravity to keep its molecules from escaping when heated by the stellar fluxes. Obviously, the Moon, which is as much in the habitable zone than the Earth, has been too small. This seems also true for Mars, in spite of the fact that it is further from the Sun and that its $CO_2$ atmosphere would have been a very good exobase radiative cooler. To first order, the size limit around the Sun may be somewhere between Mars and the Earth, although it would depend on the composition of the atmosphere. For instance Lichtenegger et al. (2010) demonstrated that if the Earth had had a nitrogen-rich terrestrial atmosphere with a present-day composition during its youth before 3.8 Gyr ago, the atmosphere would have been removed within a few million years because of the extreme EUV and solar wind conditions that are expected to have prevailed during that period when the Sun was younger. Therefore, they concluded that a $CO_2$ amount in the early nitrogen-rich terrestrial atmosphere of at least two orders of magnitude higher than the present-time level was needed to radiatively cool and confine the upper atmosphere and protect it from complete destruction. Interestingly, this was consistent with the fact that an elevated concentration of CO2 seems necessary on early Earth to compensate for the lower luminosity of the young sun and solve the "faint young Sun paradox".

A large majority of stars in our galaxy are smaller than the Sun. In the habitable zone of such stars, and in particular in an M dwarf system, atmospheric escape may be stronger because, for a given total stellar flux, the energetic radiations and the plasma flows are relatively stronger because of the stellar activity. In such conditions, to keep its atmosphere, a gravity significantly higher than on the Earth may be necessary (if sufficient), especially if the atmosphere evolves to an $N_2$-rich atmosphere early in its lifetime (see e.g. Lammer et al. 2011, Tian 2011).

However, if the planet's gravity is large enough, another problem may occur: the planet may not be able to get rid of its hydrogen-rich proto-atmosphere (see details in Lammer et al. 2011 and reference therein). In other words, with a solid body slightly more massive than the Earth, a potential "super-Earth" may ultimately remain like Neptune, with a massive $H_2$-He envelope that would prevent water from being liquid by keeping the surface pressure too high.

Intriguing observational evidence relevant to this issue are provided by the characteristics of some of the "super Earth" for which radii and masses have been measured (Lammer et al. 2011). Kepler-11b and Kepler-11f, exhibit masses of ~$4.3 M_{Earth}$ and ~$2.3 M_{Earth}$ and radii of ~$1.97 R_{Earth}$ and ~$2.61 M_{Earth}$, which result in mean densities of 3.1 and 0.7 g cm$^{-3}$ (e.g. Borucki et al. 2011; Lissauer et al. 2011). The "super-Earth" Gliese 1214b, with a radius of ~$2.68 R_{Earth}$ and a mass of ~$6.55 M_{Earth}$, corresponds to a mean density of 1.87 g cm$^{-3}$ (e.g. Charbonneau et al. 2009). These low densities indicate substantial envelopes of light gases such as H and He or possibly H₂O and H. Could it be that these "super-Earths" could not lose their initial proto-atmospheres and that they are, in fact, "mini-Neptunes" ? In fact, the only observed "super-Earths" with higher densities which indicate rocky bodies such as the CoRoT-7b (Léger et al. 2009), , Kepler-10b (Batalha et al. 2011), Kepler-18b or Kepler-20b (Borucki et al. 2011) are much closer to their star, at a distance where really strong atmospheric escape is expected.

In summary, and to oversimplify this issue, it is not impossible that the Earth have enjoyed the right size, the right kind of star, and the right upper atmosphere composition throughout its history to keep a "good" atmosphere ($10^{-1} – 10^2$ bars) for billions of years, while being able to quickly loose its thick primordial hydrogen-rich envelope. Around different stars, such as active M stars, a larger size inducing a stronger gravity may be more appropriate to retain an atmosphere suitable for liquid water and life.

## 4.2 Habitability and geologic activity

The classical theory of habitability described in section 3.3 and the current definition of the habitable zone relies on the assumption that there is long-term stabilization of the surface temperature and $CO_2$ level due to the carbonate-silicate cycle. Without this stabilization, Earth would not be habitable, and the habitable zone would be severely limited in size. On Earth, the slow increase of solar fluxes has always been compensated by a decrease in the greenhouse effect, and accidental excursions of the climate toward global glaciation (e.g., Hoffman et al. 1998) are thought to have been counterbalanced by the $CO_2$ greenhouse effect without the interruption of life.

The key process allowing the carbonate-silicate cycle on Earth, and more generally the long-term recycling of atmospheric components chemically trapped at the surface, is plate tectonics. This is a very peculiar regime induced by the convection in the mantle, which results from the geothermal heat gradient and surface cooling. How likely is the existence of plate tectonics elsewhere? Is the geophysical stabilization of the climate necessary to maintain life a rare phenomenon? In the solar system, Earth plate tectonics is unique and its origin not well understood. Other terrestrial planet or satellites are characterized by a single "rigid lid" plate surrounding the planet, and this may be the default regime on extrasolar terrestrial planets. Plate tectonics is a complicated process that primarily requires lithospheric failure, deformation and subduction (The lithosphere is the "rigid layer" forming the plates that include the crust and the uppermost mantle). To enable plate tectonics, two conditions have been suggested : 1) Mantle convective stresses large enough to overcome lithospheric resistance to allow plate braking and 2) Plates denser (i.e., colder) than the underlying asthenosphere, to drive plate subduction. On planets smaller than the Earth (e.g., Mars), the rapid interior cooling corresponds to a weak convection stress and a thick lithosphere, and no plate tectonics is expected to be maintained in the long term. On larger planets (i.e., "Super-Earths"), available studies have reached very different views. On the one hand, in their theoretical study entitled "Inevitability of Plate tectonics and Super-Earths", Valencia et al. (2007) showed that, as the planetary mass increases, convection should be more vigorous, making the lithosphere thinner (and therefore reducing lithospheric strengh), while increasing the convective stresses (owing to larger velocities in the mantle). Such conditions should lead to plate tectonics (see also Valencia and O'Connell, 2009, Van Heck and Tackley, 2011).. On the other hand, on the basis of numerical mantle convection simulations, O'Neill and Lenardic (2007) showed that increasing the planetary radius acts to decrease the ratio of convective stresses to lithospheric resistance. They concluded that super-sized Earths are likely to be in an "episodic or stagnant" lid regime rather than in a plate tectonics regime. Who is right? In fact, the thermo-tectonic evolution of terrestrial planets is a complex combination of phenomena, which has not yet been accurately modeled. For instance, most models mentioned above did not take into account the fact that in super-Earths, the very high internal pressure increases the viscosity near the core-mantle boundary, resulting in a highly "sluggish" convection regime in the lower mantles of those planets which may reduce the ability of plate tectonics (, Stein et al. 2011, Stamenkovic et al. 2012). The effect of size on plate density and subduction has not yet been studied in detail. What these studies highlight is the possibility that the Earth may be very "lucky" to be in an exact size range (within a few percent) that allows for plate tectonics. Furthermore, Venus, which is about the size of the Earth but does not exhibit plate tectonics, shows that the Earth case may be rare, and that many factors control the phenomenon. On Venus, for instance, it is thought that the mantle is drier than on Earth, and that consequently it is more viscous and the lithosphere thicker (Nimmo and McKenzie, 1998). Similar considerations led Korenaga (2010) to conclude that the likelihood of plate tectonics is also controlled largely by the presence of surface water. Plate tectonic may also strongly depend on the history and the evolution of the planet. Using their state of the art model of coupled mantle convection and planetary tectonics, Lenardic and Crowely (2012) found that multiple tectonic modes could exist for equivalent planetary parameter values, depending on the specific geologic and climatic history. Interestingly, the existence of such "multiple tectonic modes", for equivalent parameter values, points to a reason why different groups can reach different conclusions regarding the tectonic state of extrasolar terrestrial planets larger than Earth ("super-Earths"). But it also add tremendous complexity in the question of whether extrasolar terrestrial planets will have plate tectonics.

## 5 Conclusions

The ongoing exoplanet detection surveys will soon confirm the high frequency of terrestrial planet in the habitable zone. Theoretical 3D climate studies, which benefit from our experience in modelling terrestrial atmospheres in the solar system, should allow us to estimate if liquid water can be stable on the surface of these bodies, with some accuracy. However, we will still have to make assumptions on the atmospheres. Ultimately our estimation of the frequency of habitable planets and especially of worlds able to remain habitable for billions of years will depend on our understanding of the nature and of the possible evolution of the atmospheres. Our experience in the solar system is not sufficient to estimate what may happen in other stellar systems or on a planet with a different mass than the Earth. In particular, it is not impossible that the Earth, because of its exact size, location, history, as well its sun and its planetary system has enjoyed a combination of favorable conditions (see a list of additional possible problems that the Earth seems to have avoided in Lammer et al. 2009, in the book by Ward and Brownlee, 2000, and its review by Kasting 2000). Because, by definition, we conduct our research from a habitable planet, we cannot generalize our experience assuming that it is universal.

Fortunately, we can hope that in the future it will be possible to learn more about exoplanetary atmospheres .thanks to telescopic observations and spectroscopy. An important step will be achieved in the next decade by space telescopes like the James Webb Space Telescope (JWST) or ECHO (Tinetti et al. 2012), as well as by Earth-based telescopic observations using new generation telescope like the European Extremely Large Telescope. These projects will notably be able to perform atmospheric spectroscopy on exoplanets transiting in front of their star as seen from the Earth. However, characterizing atmospheres of terrestrial planets in or near the habitable zone will remain challenging. Furthermore, the number of observable planets at suitable distance will probably be very low.Nevertheless, well before the time when we will be able to detect and characterize a truly habitable planet, the first observations of terrestrial exoplanet atmospheres, whatever they show, will allow us to make a major progress in our estimation of the likeliness of life (and especially of evolved life) elsewhere in the universe.

# References

Batalha, N. M. et al. (2011). Kepler's First Rocky Planet: Kepler-10b. The Astrophysical Journal 729, 27.

Beaulieu et al. 2006, Discovery of a cool planet of 5.5 Earth masses through gravitational microlensing. Nature, 439:437-440.

Bonfils, X., Delfosse, X., Udry, S., Forveille, T., Mayor, M., Perrier, C., Bouchy, F., Gillon, M., Lovis, C., Pepe, F., Queloz, D., Santos, N. C., Segransan, D., and Bertaux, J.-L. (2013). The HARPS search for southern extra-solar planets XXXI. The M-dwarf sample. Astronomy and Astrophysics 549, A109..Borucki W. J. et al. (2011). Characteristics of Planetary Candidates Observed by Kepler. II. Analysis of the First Four Months of Data. Astrophysical Journal, 736:19.

Borucki, W. J. et al. (2012). Kepler-22b: A 2.4 Earth-radius Planet in the Habitable Zone of a Sun-like Star. The Astrophysical Journal 745, 120.

Brack, A. (1993). Liquid water and the origin of life. Orig. of Life, 3-10:23.

Buccino, A. P., Lemarchand, G. A., and Mauas, P. J. D. (2007). UV habitable zones around M stars. Icarus, 192:582-587.

Carter, B. (2008). Five- or six-step scenario for evolution? International Journal of Astrobiology 7, 177-182.

Cassan et al. (2012) One or more bound planets per Milky Way star from microlensing observations. Nature 481, 167-169.

Cernicharo, J. and Crovisier, J. (2005). Water in Space: The Water World of ISO. Space Science Reviews, 119:29 , 69.

Charbonneau, D., Berta, Z. K., Irwin, J., Burke, C. J., Nutzman, P., Buchhave, L. A., Lovis, C., Bon_ls, X., Latham, D. W., Udry, S., Murray-Clay, R. A., Holman, M. J., Falco, E. E., Winn, J. N., Queloz, D., Pepe, F., Mayor, M., Delfosse, X., and Forveille, T. (2009). A super-Earth transiting a nearby low-mass star. Nature, 462:891-894.

Ehrenfreund, P., M. Spaans and N. G.Holm (2011). The evolution of organic matter in space. Royal Society of London Philosophical Transactions Series A 369, 538-554.

Forget, F., Hourdin, F., Fournier, R., Hourdin, C., Talagrand, O., Collins, M., Lewis, S. R., Read, P. L., and Huot., J.-P. (1999). Improved general circulation models of the Martian atmosphere from the surface to above 80 km. J. Geophys. Res., 104:24,155 pp24,176.

Forget, F. and Pierrehumbert, R. T. (1997). Warming early Mars with carbon dioxide clouds that scatter infrared radiation. Science, 278:1273 -1276.

Fressin, F., Torres, G., Charbonneau, D., Bryson, S. T., Christiansen, J., Dressing, C. D., Jenkins, J. M., Walkowicz, L. M., Batalha, N. M. (2013). The False Positive Rate of Kepler and the Occurrence of Planets. The Astrophysical Journal 766, 81.

Gerard, J.-C., Hauglustaine, D. A., and Francois, L. M. (1992). The faint young sun climatic paradox: a simulation with an interactive seasonal climate-sea ice model. Paleogeogr., Paleoclimatol., Paleoecol., 97:133-150.

Henning, T & Salama, F (1998). Carbon in the Universe. Science 282, 2204-2206.

Hoffman, P. F., Kaufman, A. J., Halverson, G. P., and Schrag, D. P. (1998). A neoproterozoic snowball Earth. Science, 281:1342-1346.

Hourdin, F., Le Van, P., Forget, F., and Talagrand, O. (1993). Meteorological variability and the annual surface pressure cycle on Mars. J. Atmos. Sci., 50:3625-3640.

Hourdin, F., Talagrand, O., Sadourny, R., R_egis, C., Gautier, D., and McKay, C. P. (1995). General circulation of the atmosphere of Titan. Icarus, 117:358-374.

Howard, A. W., Marcy, G. W., Johnson, J. A., Fischer, D. A., Wright, J. T., Isaacson, H., Valenti, J. A., Anderson, J., Lin, D. N. C., and Ida, S. (2010). The Occurrence and Mass

Distribution of Close-in Super-Earths, Neptunes, and Jupiters. Science, 330:653-.

Joshi, M. (2003). Climate Model Studies of Synchronously Rotating Planets. Astrobiology, 3:415-427.

Kaltenegger L. & Selsis F (2007). Biomarkers set in context, in Extrasolar Planets., R. Dvorak (ed), Wiley-VCH, Berlin, pp 75–98.

Kaltenegger, L., Traub, W. A., Jucks, K. W (2007). Spectral Evolution of an Earth-like Planet. The Astrophysical Journal 658, 598-616.

Kaltenegger, L., Segura, A., Mohanty, S. (2011). Model Spectra of the First Potentially Habitable Super-Earth Gl581d. The Astrophysical Journal 733, 35.

Kasting, J. F. (1988). Runaway and moist greenhouse atmosphere and the evolution of Earth and Venus. Icarus, 74:472-494.

Kasting, J. F. (1997). Warming early earth and mars. Science, 276:1213-1215.

Kasting, J. F. (2001). Essay Review: P. Ward and D. Brownlee's "Rare Earth". Perspective in biology and Medecine, 44, 117-131.

Kasting, J.F., Whitmire, D. P., and Reynolds, R. T. (1993). Habitable zones around main sequence stars. Icarus, 101:108-128.

Korenaga, J. (2010). On the Likelihood of Plate Tectonics on Super-Earths: Does Size Matter?. The Astrophysical Journal 725, L43-L46.

Lammer, H., Bredeh oft, J. H., Coustenis, A., Khodachenko, M. L., Kaltenegger, L., Grasset, O., Prieur, D., Raulin, F., Ehrenfreund, P., Yamauchi, M., Wahlund, J., Grie_meier, J., Stangl, G., Cockell, C. S., Kulikov, Y. N., Grenfell, J. L., and Rauer, H. (2009). What makes a planet habitable? The Astronomy and Astrophysics Review, 17:181-249.

Lammer, H., Kislyakova, K. G., Odert, P., Leitzinger, M., Schwarz, R., Pilat-Lohinger, E., Kulikov, Y. N., Khodachenko, M. L., Güudel, M., and Hanslmeier, A. (2011). Pathways to Earth-Like Atmospheres. Extreme Ultraviolet (EUV)-Powered Escape of Hydrogen-Rich Protoatmospheres. Origins of Life and Evolution of the Biosphere, 41:503-522.

Lebonnois, S., Burgalat, J., Rannou, P., and Charnay, B. (2012). Titan global climate model: A new 3-dimensional version of the IPSL Titan GCM. Icarus, in press.

Lebonnois, S., Hourdin, F., Eymet, V., Crespin, A., Fournier, R., and Forget, F. (2010). Superrotation of Venus' atmosphere analyzed with a full general circulation model. Journal of Geophysical Research (Planets), 115(E14):E06006.

Leconte, J., Forget, F., Charnay, B., Wordsworth, R., Selsis, F., Millour, E. (2013). 3D climate modeling of close-in land planets: Circulation patterns, climate moist bistability and habitability. Astronomy.and Astrophysics, in press. ArXiv e-prints arXiv:1303.7079.

Léger, A. et al. (2009). Transiting exoplanets from the CoRoT space mission. VIII. CoRoT-7b: the first super-Earth with measured radius. Astronomy and Astrophysics 506, 287-302.

Lenardic, A. and Crowley, J. W. (2012). On the Notion of Well-defined Tectonic Regimes for Terrestrial Planets in this Solar System and Others. The Astrophysical Journal 755, 132.

Lichtenegger, H. I. M., Lammer, H., Grie_meier, J.-M., Kulikov, Y. N., von Paris, P., Hausleitner, W., Krauss, S., and Rauer, H. (2010). Aeronomical evidence for higher CO2 levels during Earth Hadean epoch. Icarus, 210:1-7.

Lissauer, J. J.et al., (2011). A closely packed system of low-mass, low-density planets transiting Kepler-11. Nature, 470:53-58.

Longdoz, B. and Francois, L. M. (1997). The faint young sun paradox: influence of the continental configuration and of the seasonal cycle on the climatic stability. Global and Planetary change, 14:97-112.

Mayor, M., X. Bonfils, T. Forveille, X. Delfosse, S. Udry, J.-L. Bertaux, H. Beust, F. Bouchy, C. Lovis, F. Pepe, C. Perrier, D. Queloz, and N. C. Santos (2009). The HARPS search for south- ern extra-solar planets. XVIII. An Earth-mass planet in the GJ 581 planetary system. Astron- omy and Astrophysics, 507:487-294. Mayor, M. and Queloz, D. (2012). From 51 Peg to Earth-type planets. New Astronomy Reviews, 56:19-24.

McKay, C. P., Porco C. C., Altheide, T., Davis, W. L., and Kral, T. A. (2008). The Possible Origin and Persistence of Life on Enceladus and Detection of Biomarkers in the Plume. Astrobiology, 8:909-919.

Nimmo, F. and McKenzie, D. (1998). Volcanism and Tectonics on Venus. Annual Review of Earth and Planetary Sciences, 26:23-53.

O'Neill, C. and Lenardic, A. (2007). Geological consequences of super-sized Earths. Geophys. Res. Lett., 34:19204-+.

Pepe, F., Lovis, C., Ségransan, D., Benz, W., Bouchy, F., Dumusque, X., Mayor, M., Queloz, D., Santos, N. C., Udry, S (2011). The HARPS search for Earth-like planets in the habitable zone. I. Very low-mass planets around HD 20794, HD 85512, and HD 192310. Astronomy and Astrophysics 534, A58.

Pierrehumbert, R. and Gaidos, E. (2011). Hydrogen Greenhouse Planets Beyond the Habitable Zone. The Astrophysical Journal Letters, 734:L13.

Rosing, M. T. (2005). Thermodynamics of life on the planetary scale. International Journal of Astrobiology, 4:9-11.

Rothschild, L. J. and Mancinelli, R. L. (2001). Life in extreme environments. Nature, 409:1092-1101.

Sagan, C. (1996). Circumstellar habitable zones: an introduction. In Doyle, L. R., editor, Circumstellar Habitable Zones, pages 3-14.

Sagan, C. and Chyba, C. (1997). The early faint young sun paradox: organic shielding of ultraviolet-labile greenhouse gases. Science, 276:1217-1221.

Santerne, A., Diaz, R. F., Moutou, C., Bouchy, F., Hébrard, G., Almenara, J.-M., Bonomo, A. S., Deleuil, M., Santos, N. C. (2012). SOPHIE velocimetry of Kepler transit candidates. VII. A false-positive rate of 35\% for Kepler close-in giant candidates.\ Astronomy and Astrophysics 545, A76.

Sellers, W. (1969). A climate model based on the energy balance of the Earth-atmosphere system. J. Appl. Met., 8:392-400.

Selsis, F., Kasting, J. F., Levrard, B., Paillet, J., Ribas, I., and Delfosse, X. (2007). Habitable planets around the star Gliese 581? Astronomy and Astrophysics, 476:1373-1387.

Spiegel, D. S., Menou, K., and Scharf, C. A. (2008). Habitable Climates. Astrophys. Jour., 681:1609-1623.

Stamenkovic, V., Noack, L., Breuer, D., and Spohn, T. (2012). The Inuence of Pressure-dependent Viscosity on the Thermal Evolution of Super-Earths. Astrophysical Journal, 748:41.

Stein, C., A. Finnenkötter, J. P. Lowman, and U. Hansen (2011), The pressure-weakening effect in super-Earths: Consequences of a decrease in lower mantle viscosity on surface dynamics, Geophys. Res. Lett., 38, L21201, doi:10.1029/2011GL049341.

Tarter, J. C. et al., (2007). A Reappraisal of The Habitability of Planets around M Dwarf Stars. Astrobiology, 7:30-65.

Tian, F. (2011). The Nitrogen Constraint on Habitability of Planets around Low Mass M-stars. In EPSC-DPS Joint Meeting 2011, page 380.

Tinetti et al. (2012). EChO. Experimental Astronomy, page 35.

Toon, O. B., McKay, C. P., Ackerman, T. P., and Santhanam, K. (1989). Rapid calculation of radiative heating rates and photodissociation rates in inhomogeneous multiple scattering atmospheres. J. Geophys. Res., 94:16,287-16,301.

Tuomi, M., Anglada-Escudé, G., Gerlach, E., Jones, H. R. A., Reiners, A., Rivera, E. J., Vogt, S. S., Butler, R. P. (2013). Habitable-zone super-Earth candidate in a six-planet system around the K2.5V star HD 40307. Astronomy and Astrophysics 549, A48.

Udry, S., X. Bonfils, X. Delfosse, T. Forveille, M. Mayor, C. Perrier, F. Bouchy, C. Lovis, F. Pepe, D. Queloz, and J.-L. Bertaux (2007). The HARPS search for southern extra-solar planets. XI. Super-Earths (5 and 8 M+) in a 3-planet system. Astron. Astrophys., 469:L43.Valencia, D., O'Connell, R. J., and Sasselov, D. D. (2007). Inevitability of Plate Tectonics onSuper-Earths. Astrophysical Journal, 670:L45-L48.

Valencia, D., O'Connell, R.~J. (2009). Convection scaling and subduction on Earth and super-Earths. Earth and Planetary Science Letters 286, 492-502.

Van Heck, H. J. and Tackley, P. J. (2011). Plate tectonics on super-Earths: Equally or more likely than on Earth. Earth and Planetary Science Letters 310, 252-261.

Von Paris, P., Gebauer, S., Godolt, M., Grenfell, J. L., Hedelt, P., Kitzmann, D., Patzer, A. B. C., Rauer, H., Stracke, B. (2010). The extrasolar planet Gliese 581d: a potentially habitable planet? Astronomy and Astrophysics 522, A23.

Von Paris, P., Gebauer, S., Godolt, M., Rauer, H., Stracke, B (2011). Atmospheric studies of habitability in the Gliese 581 system. Astronomy and Astrophysics 532, A58.

Walker, J. C. G., Hays, P. B., and Kasting, J. F. (1981). A negative feedback mechanism for the long term stabilization of the earth's surface temperature. J. Geophys. Res., 86:9776-9782.

Ward, P. D. and Brownlee, D. (2000). Rare Earth. Why complex life is uncommon in the universe. Copernicus Books.

Williams, D. M. and Kasting, J. F. (1997). Habitable planets with high obliquities. Icarus, 129:254-267.

Wordsworth, R. (2012). Transient conditions for biogenesis on low-mass exoplanets with escaping hydrogen atmospheres. Icarus, 219:267-273.

Wordsworth, R., Forget, F., and Eymet, V. (2010a). Infrared collision-induced and far-line absorption in dense $CO_2$ atmospheres. Icarus, 210:992-997.

Wordsworth, R. D., Forget, F., Selsis, F., Madeleine, J.-B., Millour, E., and Eymet, V. (2010b). Is Gliese 581d habitable? Some constraints from radiative-convective climate modeling. Astron. Astrophys., 522:A22.

Wordsworth, R. D., Forget, F., Selsis, F., Millour, E., Charnay, B., and Madeleine, J.-B. (2011). Gliese 581d is the First Discovered Terrestrial-mass Exoplanet in the Habitable Zone. The Astrophysical Journal Letters, 733:L48.

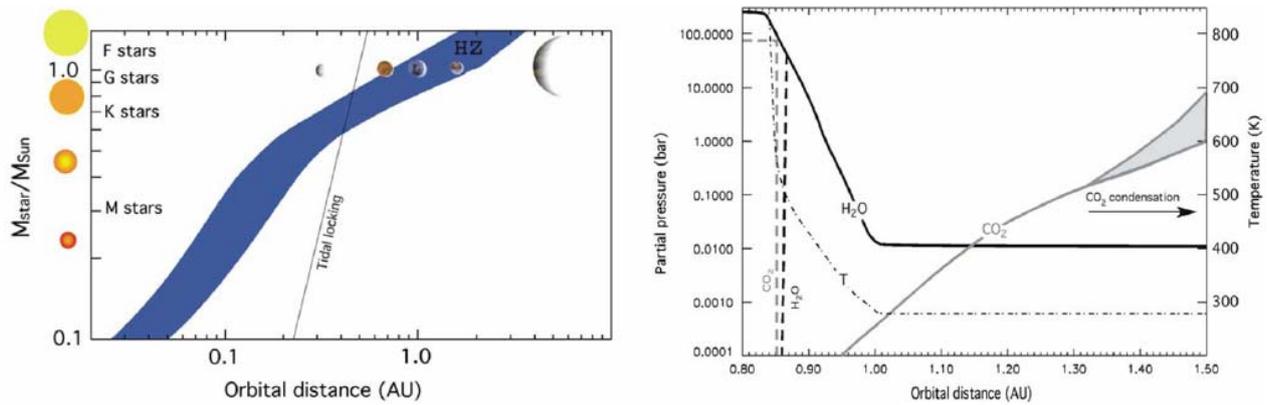

**Figure 1**. The classical habitable zone (left) and the main atmospheric composition (right) of an Earth-analog atmosphere as a function of distance from its host star. The classical habitable zone theory assumes that geophysical cycles will adjust the atmospheric $CO_2$ content and its greenhouse effect to compensate for weaker radiation flux when distant from the star. Without that, the habitable zone would be here a thin blue line. On the right, the Dashed-dotted line represents the surface temperature of the planet and the dashed lines correspond to the inner edge of the habitable zone. The grey zone correspond to the incertitude related to the effects of $CO_2$ ice clouds (see text). Figure from Kaltenegger and Selsis (2007) and Lammer et al. (2009) with data from Kasting et al. (1993) and Forget and Pierrehumbert (1997).

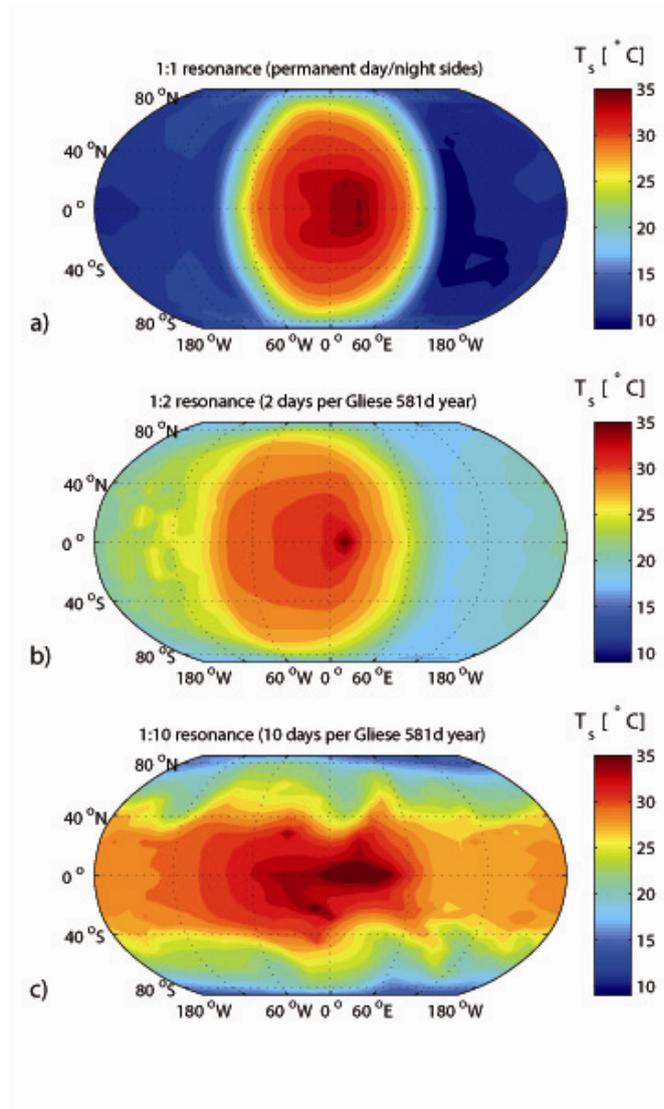

**Figure 2.** Surface temperature snapshots from 3D global climate model simulations for extrasolar planet Gliese 581d, assuming a 20-bar $CO_2$ atmosphere and for three possible rotation rates. Such 3D simulations can help better understand the habitability of exoplanets, although it is necessary to make strong assumptions on the nature of the atmosphere. Figure from Wordsworth et al. 2011.